\documentclass[pra,aps,preprint,showpacs]{revtex4}
\usepackage{epsfig,amssymb,amsfonts}

\def\opone{\leavevmode\hbox{\small1\kern-3.8pt\normalsize1}}

\begin{document}

\title{Effects of anisotropy on optimal dense coding}
\author{Guo-Feng Zhang\footnote{Corresponding author.}\footnote{Email:
gf1978zhang@buaa.edu.cn}} \affiliation{Department of physics, School
of sciences, Beijing University of Aeronautics and Astronautics,
Xueyuan Road No. 37, Beijing 100083, People's Republic of China }

\begin{abstract}
We study optimal dense coding with thermal entangled states of a
two-qubit anisotropic \emph{XXZ} model and a Heisenberg model with
Dzyaloshinski-Moriya (DM) interactions. The DM interaction is
another kind of anisotropic antisymmetric exchange interaction. The
effects of these two kinds of anisotropies on dense coding are
studied in detail for both the antiferromagnetic and ferromagnetic
cases. For the two models, we give the conditions that the
parameters of the models have to satisfy for a valid dense coding.
We also found that even though there is entanglement, it is
unavailable for our optimal dense coding, which is the same as
entanglement teleportation.
\end{abstract}


\maketitle

\section{Introduction}

Entanglement is one of the most fascinating features of quantum
mechanics and plays a central role in quantum information
processing, such as quantum key distribution\cite{C. H. Bennett1},
quantum teleportation \cite{A. K. Ekert}, dense coding \cite{C. H.
Bennett2}, and so on. In the initial dense coding protocol \cite{C.
H. Bennett3}, the sender can transmit two bits of classical
information to the receiver by sending a single qubit if they share
a two-qubit maximally entangled state (an Einstein-Podolsky-Rosen
(EPR) state). Since then, many works on dense coding have been
presented experimentally \cite{K. Mattle} or theoretically \cite{A.
Bareno, S. L. Braunstein, S. Bose}. We know, in a general dense
coding, the sender performs one of the local unitary transformations
$U_{i}$$\in U(d)$ on $d$-dimensional quantum system to put the
initially shared entangled state $\rho$ in $\rho_{i}=(U_{i}\otimes
I_{d})\rho(U_{i}^{\dagger}\otimes I_{d})$ with a priori probability
$p_{i}$($i=0,1,...,i_{max}$), and then the sender sends off his
quantum system to the receiver. Upon receiving this quantum system,
the receiver performs a suitable measurement on $\rho_{i}$ to
extract the signal. The optimal amount of information that can be
conveyed is known to be bounded from by the Holevo quantity\cite{A.
S. Holevo}:
$\chi=S(\overline{\rho})-\sum_{i=0}^{i_{max}}p_{i}S(\rho_{i})$,
where $S(\rho)$ denotes the von Neumann entropy and
$S(\overline{\rho})=\sum_{i=0}^{i_{max}}p_{i}\rho_{i}$ is the
average density matrix of the signal ensemble. Since the Holevo
quantity is asymptotically achievable\cite{A. S. Holevo1}, one can
use $\chi=S(\overline{\rho})-\sum_{i=0}^{i_{max}}p_{i}S(\rho_{i})$
as the definition of the capacity of dense coding. Moreover, the von
Neumann entropy is invariant under unitary transformations,
$S(\rho_{i})=S(\rho)$. Therefore, the dense coding capacity can be
rewritten as $\chi=S(\overline{\rho})-S(\rho)$. The following
problem is to find the optimal signal ensemble
$\{\rho_{i};p_{i}\}_{i=0}^{i_{max}}$ that maximizes $\chi$. In
Ref.\cite{T. Hiroshima}, the author showed that the $d^{2}$ signal
states ($i_{max}=d^{2}-1$) generated by mutually orthogonal unitary
transformations with equal probabilities yield the maximum $\chi$,
which is called optimal dense coding, and considered the optimal
dense coding when the shared entangled state was a general mixed
one.

The quantum entanglement in solid state systems such as spin chains
has been an important emerging field since the founding of thermal
entanglement \cite{M. A. Nielsen}. Spin chains are natural
candidates for the realization of the entanglement compared with
other physical systems. As the thermal fluctuation is introduced
into the system, the state of a typical solid-state system at
thermal equilibrium (temperature $T$) is
$\rho(T)=e^{-\beta\emph{H}}/Z$, where $H$ is the Hamiltonian,
$Z=tre^{-\beta\emph{H}}$ is the partition function and
$\beta=1/(k\emph{T})$, where $k$ is the Boltzman constant. For
simplicity, we write $k=1$. As $\rho(T)$ represents a thermal state,
the entanglement in the state is called the thermal entanglement.
The study of thermal entanglement properties in Heisenberg systems
has received a great deal of  attention \cite{X. Wang, D. V.
Khveshchenko, U. Glaser, G. Vidal, G. K. Brennen, F. Verstraete}.
Some authors have considered the availability of thermal
entanglement. Quantum teleportation that uses the thermal entangled
state as a channel has been proposed in Refs. \cite{Y. Yeo, Z. C.
Kao, G. F. Zhang2}.

In this paper, we study the optimal dense coding \cite{T. Hiroshima}
using the thermal entangled states of a two-qubit anisotropic XXZ
model and a Heisenberg model with DM interactions. We investigate
the effects of two kinds of anisotropies on dense coding in detail
for both the antiferromagnetic (AFM) and ferromagnetic (FM) cases
and give the conditions that the parameters of the model have to
satisfy for dense coding. The paper is organized as follows: in Sec.
II, the optimal dense coding using the thermal entangled states of a
two-qubit anisotropic \emph{XXZ} model is investigated; we study the
optimal dense coding using the thermal entangled state of a
Heisenberg model with DM interactions in Sec. III and conclude in
Sec. IV.

\section{Optimal dense coding using thermal states of a two-qubit anisotropic \emph{XXZ} chain}

We consider a two-qubit anisotropic \emph{XXZ} Heisenberg model
\begin{equation}
H=\frac{J}{2}(\sigma_{1x}\sigma_{2x}+\sigma_{1y}\sigma_{2y}+\Delta\sigma_{1z}\sigma_{2z})=J(\sigma_{1+}\sigma_{2-}+\sigma_{1-}\sigma_{2+})+\frac{J
\Delta}{2}\sigma_{1z}\sigma_{2z},
\end{equation}
where $\sigma_{j\alpha}$ $(j=1,2, \alpha=x, y, z)$ are the pauli
matrices. $J$ is the real coupling constant, $J > 0$ corresponding
to the antiferromagnetic (AFM) case and $J < 0$ to the ferromagnetic
(FM) case. The operators $\sigma_{j\pm} = (1/2)(\sigma_{jx}\pm
i\sigma_{jy})$. Without loss of generality, we define $|0\rangle$
$(|1\rangle)$ as the ground (excited) state of a two-level particle.
The eigensystem of $H$ is $H|00\rangle=\frac{J\Delta}{2}|00\rangle$,
$H|\Psi^{\pm}\rangle=(-J\Delta/2\pm J)|\Psi^{\pm}\rangle$ and
$H|11\rangle=\frac{J\Delta}{2}|11\rangle$, where
$|\Psi^{\pm}\rangle=(1/\sqrt{2})(|01\rangle\pm|10\rangle)$ are the
two-qubit maximally entangled states (EPR states). The thermal state
of the system at equilibrium (temperature $T$) is
\begin{equation}
\rho=\frac{1}{Z_{1}}[e^{-\beta\frac{J\Delta}{2}}|00\rangle\langle00|+e^{-\beta(-\frac{J\Delta}{2}+J)}|\Psi^{+}\rangle\langle\Psi^{+}|+e^{-\beta(-\frac{J\Delta}{2}-J)}|\Psi^{-}\rangle\langle\Psi^{-}|+
e^{-\beta\frac{J\Delta}{2}}|11\rangle\langle11|],
\end{equation}
where $Z_{1}=2\lambda e^{-J/2T}$ is the partition function and
$\lambda=1+e^{J\Delta/T}\cosh[J/T]$. In Ref. \cite{X. Wang1}, the
concurrence \cite{W. K. Wootters} of the model is considered as a
measure of thermal entanglement.

Now we carry out the optimal dense coding with the thermal entangled
states of the two-qubit system as a channel. The set of mutually
orthogonal unitary transformations \cite{T. Hiroshima} of the
optimal dense coding for two-qubit is
\begin{eqnarray}
U_{00}|x\rangle&=&|x\rangle,
U_{10}|x\rangle=e^{\sqrt{-1}(2\pi/2)x}|x\rangle,\nonumber
\\U_{01}|x\rangle&=&|x+1(\texttt{mod2})\rangle,
U_{11}|x\rangle=e^{\sqrt{-1}(2\pi/2)x}|x+1(\texttt{mod2})\rangle,
\end{eqnarray}
where $|x\rangle$ is the single qubit computational basis
$(|x\rangle=|0\rangle, |1\rangle)$. The average state of the
ensemble of signal states generated by the unitary transformations
Eq. (3) is
\begin{equation}
\overline{\rho^{*}}=\frac{1}{4}\sum^{3}_{i=0}(U_{i}\otimes
I_{2})\rho (U_{i}^{+}\otimes I_{2}),
\end{equation}
where we have assumed $0\rightarrow00$; $1\rightarrow01$;
$2\rightarrow10$; $3\rightarrow11$, and $\rho$ is the thermal states
of Eq. (2). Through straightforward algebra, we have
\begin{equation}
\overline{\rho^{*}}=\frac{1}{4}[|00\rangle\langle00|+|01\rangle\langle01|+|10\rangle\langle10|+|11\rangle\langle11|].
\end{equation}
After completing the set of mutually orthogonal unitary
transformations, the maximum dense coding capacity $\chi$ can be
written as
\begin{equation}
\chi=S(\overline{\rho^{*}})-S(\rho)=2-S(\rho).
\end{equation}
Here, $S(\rho)$ is the von Neumann entropy of the quantum state
$\rho$. Thus, the value of the maximal dense coding capacity is
\begin{equation}
\chi=\frac{T\lambda(\ln[4]-2\ln[\lambda])+2J\xi e^{\frac{J\Delta
}{T}}}{T\lambda \ln[4]},
\end{equation}
where $\xi=\Delta\cosh[\frac{J}{T}]+\sinh[\frac{J}{T}]$.
\begin{figure}
\begin{center}
\epsfig{figure=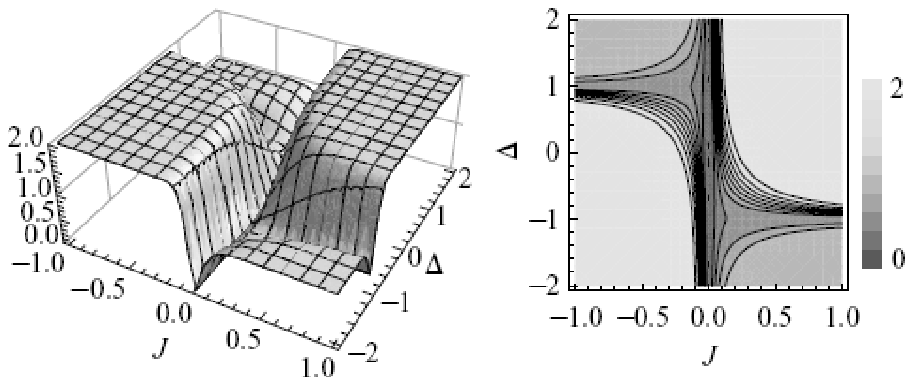, width=0.60\textwidth}
\end{center}
\caption{(Color online) The optimal dense coding capacity $\chi$
versus coupling constant $J$ and anisotropy $\Delta$. We assume
$T=0.05$. The right panel is the contour.}
\end{figure}

It is found that $\chi(J, \Delta)=\chi(-J, -\Delta)$, which
indicates that the maximal dense coding capacity $\chi$ satisfy
$\chi_{AFM}(\Delta)=\chi_{FM}(-\Delta)$.  The result is the same as
the concurrence \cite{X. Wang1}. We give the numerical analysis of
$\chi$. In Fig. 1, the optimal dense coding capacity as a function
of the coupling constant $J$ and anisotropy $\Delta$ is plotted at a
definite temperature. From the analytical point of view, in order to
carry out the optimal dense coding successfully, the parameters of
the model must satisfy
\begin{equation}
\chi>\log_{2}[2]=1\Leftrightarrow J \xi
e^{\frac{J\Delta}{T}}>T\lambda\ln[\lambda].
\end{equation}
For different values of the anisotropy $\Delta$ and the coupling
constant $J$, there must be a critical temperature $T_{critical}$ ,
beyond which we cannot give an optimal dense coding with this
two-qubit Heisenberg $\emph{XXZ}$ chain. In the following, we will
investigate explicitly the effects of $\Delta$ and $T$ on $\chi$:

Case1: Anisotropy $\Delta=0$. We find $\chi=2$ when $T\rightarrow0$,
is always true regardless of AFM or FM case. The result is same when
the channel is a two-qubit EPR state. It is because in this case,
the thermal state is $|\Psi^{-}\rangle$ (for AFM) or
$|\Psi^{+}\rangle$ (for FM), which are the EPR states, so the sender
can transmit 2 bits classical information by sending 1 qubit.
\begin{figure}
\begin{center}
\epsfig{figure=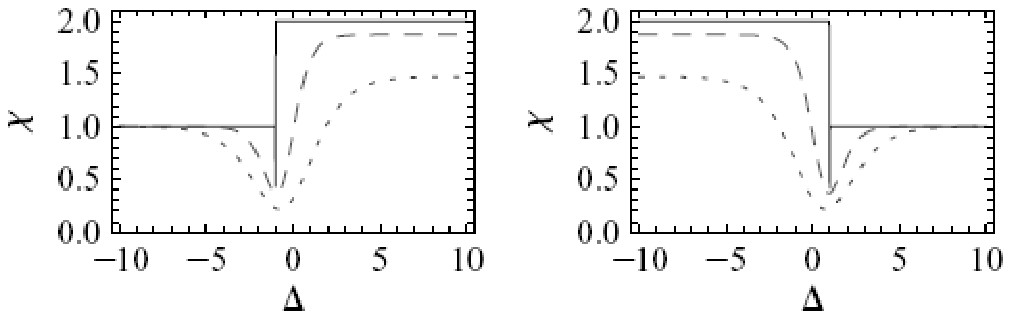}
\end{center}
\caption{(Color online) The optimal dense coding capacity $\chi$
versus anisotropy $\Delta$. The left panel corresponds to an AFM
case $(J=1)$ and the right panel corresponds to a FM one $(J=-1)$.
From top to bottom, the temperature is 0.005, 0.5, 1, respectively.}
\end{figure}

Case2: Anisotropy $|\Delta|\gg0$. We have
\begin{eqnarray}
\left\{%
\begin{array}{ll}
    \chi=\frac{1+T\ln[4]-T\ln[1+e^{2/T}]+\tanh[\frac{1}{T}]}{T\ln[2]}, & \hbox{if $\Delta\rightarrow+\infty$;} \\
    \chi=1, & \hbox{if $\Delta\rightarrow-\infty$.} \\
\end{array}%
\right.
\end{eqnarray}
for $J=1$,
\begin{eqnarray}
\left\{%
\begin{array}{ll}
    \chi=1, & \hbox{if $\Delta\rightarrow+\infty$;} \\
    \chi=\frac{1+T\ln[4]-T\ln[1+e^{2/T}]+\tanh[\frac{1}{T}]}{T\ln[2]}, & \hbox{if $\Delta\rightarrow-\infty$.} \\
\end{array}%
\right.
\end{eqnarray}
for $J=-1$. These features can be manifested from Fig.2. For $J=1$
and $\Delta\rightarrow-\infty$ or $J=-1$ and
$\Delta\rightarrow+\infty$, we have $\chi=1$, which means the
quantum channel is not valid for optimal dense coding, for now the
thermal state is
$\frac{1}{2}(|00\rangle\langle00|+|11\rangle\langle11|)$, which is a
superposition of two product states. But for $J=1$ and
$\Delta\rightarrow+\infty$ or $J=-1$ and $\Delta\rightarrow-\infty$,
the value of the maximal dense coding capacity is the same and
depends on the temperature. In these two cases, the thermal state is
\begin{eqnarray}
\rho=\frac{1}{2}
\left(%
\begin{array}{cccc}
  0& 0  &0 & 0 \\
  0 & 1 & \pm\tanh[\frac{1}{T}]& 0 \\
  0 & \pm\tanh[\frac{1}{T}] & 1 & 0\\
 0 &0 & 0 & 0 \\
\end{array}%
\right),
\end{eqnarray}
where `$+$' corresponds $J=-1$ and $\Delta\rightarrow-\infty$, `$-$'
corresponds $J=1$ and $\Delta\rightarrow+\infty$. The concurrence
\cite{W. K. Wootters} of Eq.(11) is $C=\tanh[\frac{1}{T}]$, which
means Eq. (11) is an entangled state for a finite temperature. In
order to get a valid optimal dense coding when anisotropy
$|\Delta|\gg0$, here we must have
$\frac{1+T\ln[4]-T\ln[1+e^{2/T}]+\tanh[\frac{1}{T}]}{T\ln[2]}>1\Leftrightarrow\frac{1+\tanh[1/T]}{\ln[(1+e^{2/T})/2]}>T$.
This can be held for any temperature. We can also see this from
Fig.2. Moreover, since $\chi(J=1, \Delta=-1)=\chi(J=-1,
\Delta=1)=\ln[4/3]/\ln[2]\approx0.415037$ when $T\rightarrow0$, the
value will be smaller as the temperature increases. This can be
understood since the concurrence of this model is zero when
$\Delta=-1$ for AFM and $\Delta=1$ for FM. Therefore, $\chi$ has an
abrupt transition at the two points.
\begin{figure}
\begin{center}
\epsfig{figure=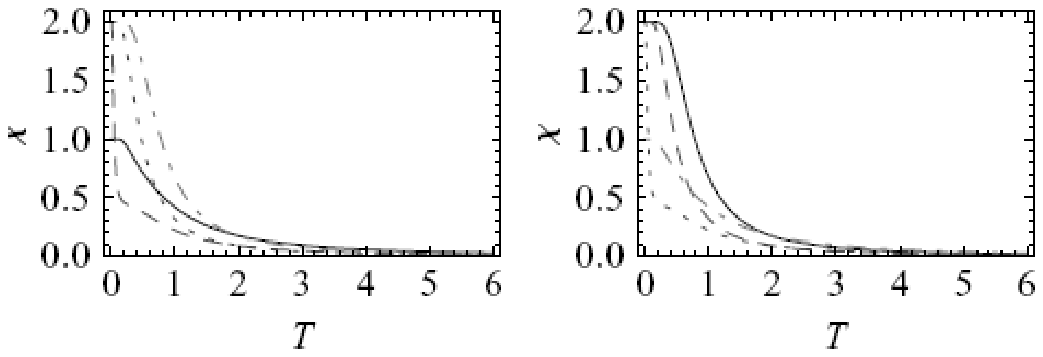}
\end{center}
\caption{(Color online) The optimal dense coding capacity $\chi$
versus temperature $T$. The left panel corresponds to an AFM case
$(J=1)$ and the right panel corresponds to a FM one $(J=-1)$. Left
panel: Solid (Black) curve for $\Delta=-2$, Dashed (Blue) curve for
$\Delta=-0.9$, Dotted (Red) curve for $\Delta=0$, Dash-dotted
(Green) curve for $\Delta=1$. Right panel: Solid (Black) curve for
$\Delta=-1$, Dashed (Blue) curve for $\Delta=0$, Dotted (Red) curve
for $\Delta=0.9$, Dash-dotted (Green) curve for $\Delta=2$.}
\end{figure}

Case3: For a finite anisotropy and temperature. we plot the optimal
dense coding capacity $\chi$ as a function of the temperature $T$
for four different values of anisotropy in Fig.3. From the figure,
when $J=1$, if $\Delta<-1$, regardless of what temperature, $\chi$
is always less than 1 and the thermal entangled states are not valid
for optimal dense coding. This can be explained since the
concurrence of this model $C_{AFM}=0$ for $\Delta<-1$. Moreover, we
can see with the increase of the $\Delta$, the area of $T$ that the
optimal dense coding is feasible becomes wider. Accordingly, for
$J=-1$, we must have $\Delta<1$ in order to make $\chi>1$ at some
temperature. This is easily understood because $C_{FM}=0$ for
$\Delta>1$. But the area of $T$ that the optimal dense coding is
feasible becomes more narrow.

\section{The effects of DM interaction on optimal dense coding}

Another kind of anisotropy that we investigate is the DM anisotropic
antisymmetric interaction which arises from spin-orbit coupling
\cite{I. Dzyaloshinskii, T. Moriya}. Now we consider the Heisenberg
model with DM interaction
\begin{equation}
\label{1}
H_{DM}=\frac{J}{2}[(\sigma_{1x}\sigma_{2x}+\sigma_{1y}\sigma_{2y}+\sigma_{1z}\sigma_{2z})+\overrightarrow{D}\cdot(\overrightarrow{\sigma_{1}}\times\overrightarrow{\sigma_{2}})],
\end{equation}
here $\overrightarrow{D}$ is the DM vector coupling. For simplicity,
we choose $\overrightarrow{D}=D\overrightarrow{z}$, and the
Hamiltonian $H_{DM}$ becomes
\begin{eqnarray}
\label{2}
H_{DM}&=&\frac{J}{2}[\sigma_{1x}\sigma_{2x}+\sigma_{1y}\sigma_{2y}+\sigma_{1z}\sigma_{2z}+D(\sigma_{1x}\sigma_{2y}-\sigma_{1y}\sigma_{2x})]
\nonumber
\\&=&J[(1+iD)\sigma_{1+}\sigma_{2-}+(1-iD)\sigma_{1-}\sigma_{2+}+\frac{J}{2}\sigma_{1z}\sigma_{2z}].
\end{eqnarray}
We notice that $H_{DM}(D=0)=H(\Delta=1)$. The eigenvalues and
eigenvectors of $H_{DM}$ are
$H_{DM}|00\rangle=\frac{J}{2}|00\rangle$,
$H_{DM}|11\rangle=\frac{J}{2}|11\rangle$, $H_{DM}|\pm\rangle=(\pm
J\sqrt{1+D^{2}}-\frac{J}{2})|\pm\rangle$, with
$|\pm\rangle=(1/\sqrt{2})(|01\rangle\pm e^{i\theta}|10\rangle)$ and
$\theta=\arctan D$.

As the thermal fluctuation is introduced into the system, in the
standard basis $\{|11\rangle,|10\rangle,|01\rangle,|00\rangle\}$,
the state can be expressed as
\begin{eqnarray}
\rho_{DM}=\frac{1}{Z_{2}}
\left(%
\begin{array}{cccc}
  e^{-\beta J/2}& 0  &0 & 0 \\
  0 & \frac{1}{2}e^{\frac{1}{2}\beta (J-\delta)}(1+e^{\beta \delta}) & \frac{1}{2}e^{i\theta}e^{\frac{1}{2}\beta (J-\delta)}(1-e^{\beta \delta}) & 0 \\
  0 &   \frac{1}{2}e^{-i\theta}e^{\frac{1}{2}\beta (J-\delta)}(1-e^{\beta \delta}) & \frac{1}{2}e^{\frac{1}{2}\beta (J-\delta)}(1+e^{\beta \delta})  & 0\\
 0 &0 & 0 & e^{-\beta J/2} \\
\end{array}%
\right),
\end{eqnarray}
where $Z_{2}=2\eta e^{-\frac{J}{2T}}$,
$\eta=1+e^{J/T}\cosh[\delta/(2T)]$ and $\delta=2J\sqrt{1+D^{2}}$.
The entanglement of this model has been studied \cite{G. F. Zhang2}
by means of concurrence. Through Eq.(3) and Eq.(4), we have also
\begin{equation}
\overline{\rho^{*}}_{DM}=\frac{1}{4}[|00\rangle\langle00|+|01\rangle\langle01|+|10\rangle\langle10|+|11\rangle\langle11|].
\end{equation}
\begin{figure}
\begin{center}
\epsfig{figure=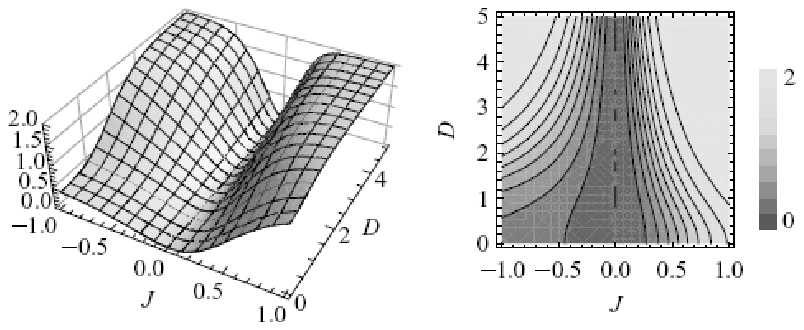,width=0.60\textwidth}
\end{center}
\caption{(Color online) The optimal dense coding capacity $\chi$
versus coupling constant $J$ and DM interaction $D$. We assume
$T=0.5$. The right panel is the contour.}
\end{figure}
Similarly, after a straight calculation, the value of the maximal
dense coding capacity is given by
\begin{equation}
\chi=\frac{T\eta(\ln[4]-2\ln[\eta])+\zeta
e^{\frac{J}{T}}}{T\eta\ln[4]},
\end{equation}
where $\zeta=2J\cosh[\delta/(2T)]+\delta\sinh[\delta/(2T)]]$. In
order to carry out the optimal dense coding successfully, the
parameters of the model must satisfy
$\chi>\log_{2}[2]=1\Leftrightarrow \zeta
e^{\frac{J}{T}}>2T\eta\ln[\eta]$, which can be returned to Eq.(8)
where $\Delta=1$ for a vanishing DM interaction.

We give a plot of optimal dense coding capacity as a function of DM
interaction and spin coupling constant at a definite temperature in
Fig.4. The variation of $\chi$ with $D$ is very similar to that of
concurrence for the AFM case. But for the FM case, the behavior of
$\chi$ is different from that of concurrence. These results can be
found by comparing Fig.4 with Fig.1 in Ref. \cite{G. F. Zhang2}.
Next, we consider the effects of DM interaction on dense coding
capacity.
\begin{figure}
\begin{center}
\epsfig{figure=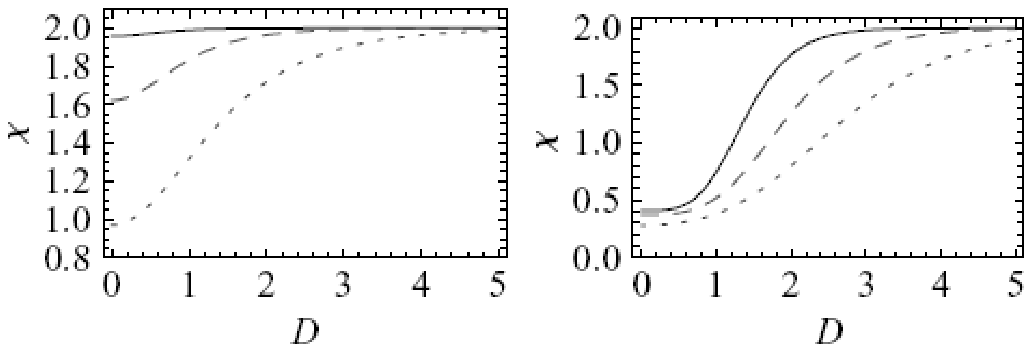}
\end{center}
\caption{(Color online) The optimal dense coding capacity $\chi$
versus DM interaction $D$. The left panel corresponds to an AFM case
$(J=1)$ and the right panel corresponds to a FM case $(J=-1)$. From
top to bottom, the temperature is 0.3, 0.5, 0.8, respectively.}
\end{figure}
\begin{figure}
\begin{center}
\epsfig{figure=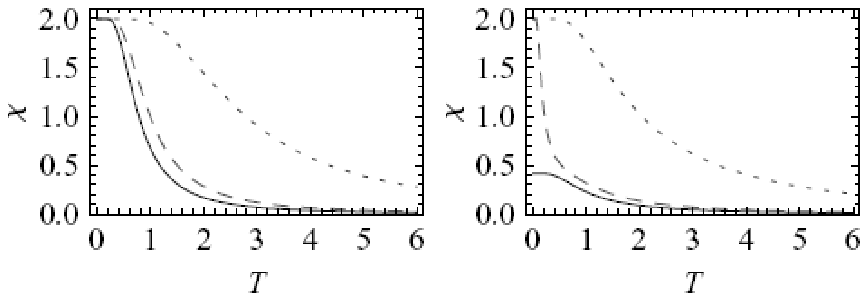}
\end{center}
\caption{(Color online) The optimal dense coding capacity $\chi$
versus the temperature. The left panel corresponds to an AFM case
$(J=1)$ and the right panel corresponds to a FM case $(J=-1)$. From
top to bottom, the DM interaction is 5, 1, 0, respectively.}
\end{figure}

Case1: DM interaction $D\gg0$. The optimal dense coding capacity is
always equal to $log_{2}[4]=2$ for the AFM and FM cases since the
thermal is $\frac{1}{2}[|01\rangle\langle01|\mp
i|01\rangle\langle10|\pm
i|10\rangle\langle01|+|10\rangle\langle10|]$, which is a EPR type
state and its concurrence is 1.

Case2: For a finite DM interaction and temperature. The variation of
$\chi$ with $D$ for $J=1$ and $J=-1$ is plotted in Fig.5. As
temperature increases, the useful area of $D$ for optimal dense
coding becomes narrow whether $J>0$ or $J<0$. Although there exists
some $D$ for the thermal state that is not valid for optimal dense
coding, for these $D$ the concurrence of the model is not zero.
Therefore, even though there is entanglement, it is unavailable for
our optimal dense coding, which is the same as entanglement
teleportation. Comparing the left with the right, DM interaction
must be stronger for FM in order to make $\chi>1$ at the same
temperature for AFM. Moreover, $D=0$, $\chi$ is always less than 2
for the FM case no matter what temperature is, which can be easily
understood since the entanglement is zero. In Fig.6, the optimal
dense coding capacity as a function of the temperature is plotted
for different DM interactions. The critical value of $T$ when the
thermal entangled states is valid for optimal dense coding
($\chi>1$) for $D=5$ is larger than that for $D=1$. At zero
temperature, regardless of the DM interaction, $\chi=2$ for the AFM
because the state is $|-\rangle=\frac{1}{\sqrt{2}}(|01\rangle-
e^{i\theta}|10\rangle)$, which is a EPR type state. However, the
thermal state is uncertain for the FM case at zero temperature. For
nonzero $D$, the state is $|+\rangle=\frac{1}{\sqrt{2}}(|01\rangle+
e^{i\theta}|10\rangle)$, so $\chi(D\neq0)=2$. At zero temperature,
$\chi(D=0)<1$ because the state is
$\frac{1}{6}[2|00\rangle\langle00|+|01\rangle\langle01|+i|01\rangle\langle10|-i|10\rangle\langle01|+|10\rangle\langle10|+2|11\rangle\langle11|]$,
which is not an entangled state.

\section{Conclusions}
In conclusion, we studied analytically the effects of two kinds of
anisotropy on the optimal dense coding in the anisotropic \emph{XXZ}
model and the Heisenberg model with DM interaction. We demonstrated
that whether the optimal dense coding is valid or not depends on
both the anisotropic parameters and the sign of exchange constants
$J$. The conditions for a valid optimal dense coding have been
given. For the AFM \emph{XXZ} model, anisotropy must be larger than
-1 and the critical temperature above which the optimal dense coding
capacity $\chi<1$ will increase as anisotropy increases. But
anisotropy must be less than 1 and the critical temperature will
decrease with the increasing of anisotropy for the FM case. The
dependence trends of optimal dense coding capacity on DM interaction
are the same for both the AFM and FM Heisenberg model with DM
interaction, and the relatively stronger DM interaction will be
helpful for optimal dense coding. We also found that even though
there is entanglement, it is unavailable for our optimal dense
coding, which is the same as entanglement teleportation.

\section{acknowledgements}
This work was supported by the National Natural Science Foundation
of China (Grant No. 10604053 and No. 10874013).

\end{document}